# Enhanced photoluminescence from self-organized rubrene single crystal surface structures

R. J. Stöhr,[1,2] G. J. Beirne,[3,4] P. Michler,[2,3] R. Scholz,[5] J. Wrachtrup,[1,2] and J. Pflaum[6]

[1] *3. Physikalisches Institut, Universität Stuttgart, 70550 Stuttgart, Germany*
[2] *Stuttgart Research Center of Photonic Engineering (SCoPE), 70569 Stuttgart, Germany*
[3] *IHFG, Universität Stuttgart, 70569 Stuttgart, Germany*
[4] *Cavendish Laboratory, Cambridge, CB3 0HE, United Kingdom*
[5] *Walter Schottky Institut, TU München, 85748 Garching, Germany*
[6] *Experimentelle Physik VI, Universität Würzburg and ZAE Bayern, 97074 Würzburg, Germany*

(Dated: 17 June 2010)

We report on crystalline pyramidal structures grown via self-organization on the rubrene (001) surface. The analysis of their spectral response by means of photoluminescence with micrometer lateral resolution reveals an intensity enhancement on-top of the surface structures. As we demonstrate this intensity increase can be related to the excitation processes at the molecular level in combination with exciton confinement within the pyramids.

The optical properties of organic single crystals have been the subject of intense research for many decades. Their low symmetry and the correlation between their structural and optical properties result in complex phenomena such as Davydov-splitting, large hyperpolarizabilities and highly anisotropic optical parameters[1–3]. In contrast to their inorganic counterparts, the optical surface properties are not affected e.g. by dangling bonds. Furthermore, exciton binding energies of up to 1 eV favor the use of ordered molecular structures for optoelectronic applications at room temperature[4,5]. However, fundamental aspects of the exciton formation, excitonic states and lifetimes are still under discussion. The existence of a delocalized excitonic band versus the possible formation of a so-called "self-trapped" excitonic state, caused by the relaxation of the local molecular environment, highlights one of these key questions[6,7]. In this context, the material dependent diffusion constants for the various excitonic species can only be estimated by laborious methods such as transient optical gratings on molecular crystals of high quality[8,9].

In this paper we describe an alternative approach to access the optical properties of ordered organic materials by analyzing the exciton dynamics in self-organized pyramidal structures on the (001) rubrene crystal surface. Their regular arrangement and crystallinity prove to be beneficial for optical studies in contrast to approaches based on randomly oriented micro- and nanocrystals grown by precipitation[10]. The structural quality and the spatial extension enable the investigation of their optical characteristics in comparison with those of the planar (001) rubrene surface by photoluminescence with micrometer spatial resolution ($\mu$-PL).

High quality rubrene crystals were grown by sublimation under streaming $H_2$. By Laue diffraction the crystal surface was identified as (001). Additional polarization studies corroborate that the microstructures are optically active indicating their crystallinity. The pyramidal baseline ratio of 2:1 is in agreement with the ratio of the unit cell vectors of the (ab)-plane. The angles between the facets and the pyramidal base plane indicate that the facets coincide with low index planes such as {103} and {013}. Details of the underlying growth mechanisms leading to these structures will be presented in a forthcoming article. $\mu$-PL measurements were performed in a helium-flow cryostat that allows cooling to 4 K. The excitation wave-length of 380 nm was realized by a frequency-doubled Ti:Sapphire laser providing 120 fs pulses at a repetition rate of 76 MHz. The laser emission was focused to a diameter of 1 μm. The PL signal was spectrally resolved by a 0.75 m spectrometer and recorded with a cooled CCD camera. Time-resolved experiments were performed by means of a fast APD with a time resolution of 40 ps.

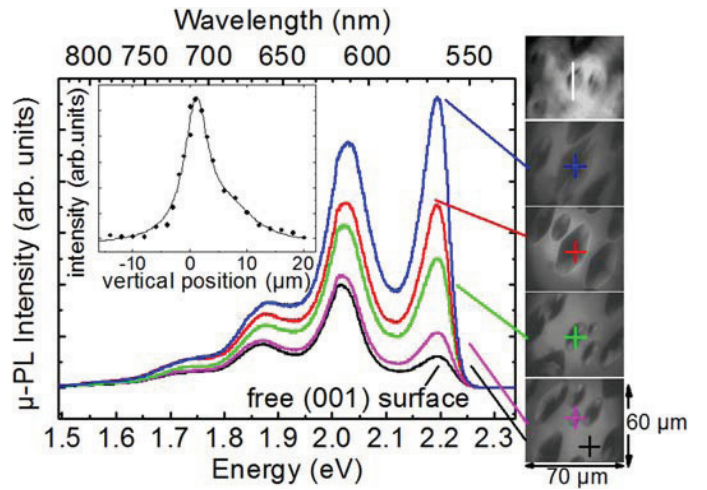

FIG. 1. PL signal recorded at 4 K from surface pyramids of various spatial extensions. The peaks can be attributed to a free exciton (2.19 eV) and a self-trapped exciton (2.02 eV) together with its vibronic replica (1.87 eV and 1.72 eV). The inset shows the integral PL intensity recorded for a line scan across a pyramidal structure on the rubrene (001) surface (white line).

The μ-PL signals recorded from several microstructures with different sizes are shown in Fig. 1. In the luminescence spectra an intensity enhancement is observed for all pyramidal structures with respect to the free rubrene surface. Comparing the intensity evolution of the 2.19 eV peak with that of the three peaks at lower energies a different dependence on the pyramidal size can be assessed including a pronounced shift in the relative intensity towards the 2.19 eV emission for larger structure sizes. To highlight this trend, Fig. 2 shows the ratio of the integrated intensities measured on the pyramids (ON) versus that on the flat surface (OFF) for all four peaks. Evidently, the 2.19 eV band is enhanced by almost one order of magnitude with increasing structure size whereas the remaining three transitions appear to be only weakly affected. Furthermore, a critical structure size can be determined above which the integrated PL intensity is almost constant for all emission bands. We exclude effects by reabsorption as these are contradictory to the observed size-dependence of the PL intensity at 2.19 eV and as reabsorption should be strongly reduced at cryogenic temperatures by the smaller spectral overlap. At first, the similar intensity enhancement with increasing structure size and the equidistant energy spacing suggest that the second set of peaks (2.02 eV, 1.87 eV, 1.72 eV) belongs to the same optical transition and its vibronic sidebands. The vibronic progression of the 2.02 eV peak (0-0 transition) can be related to the 1210 cm$^{-1}$ mode (0.15 eV) measured by Raman spectroscopy on rubrene single crystals[11]. However, to elucidate the strong enhancement of the total PL intensity from the microstructures a more sophisticated approach is needed to explain the underlying excitation processes. This again leads us to the assumption that the different emission bands may in fact not belong to the same vibronic structure but rather originate from two different excitonic species. In particular, the 2.19 eV excitation is assigned to the formation of free excitons whereas the transition at 2.02 eV is associated to self-trapped excitons[12]. This idea is supported by luminescence studies on rubrene crystals and molecules in polymeric matrices, that show an intensity shift towards the 2.02 eV emission with increasing pressure[13].

Since the detection area corresponds to the PL excitation diameter of 1 μm, excitons diffusing out of this focus will not contribute to the measured intensities. Using this constraint we are able to propose a possible scenario to rationalize the observed intensity enhancement under confined geometry as well as the resulting temperature dependence. If the exciton lifetime exceeds a certain value and, as a result, a certain diffusion length, the generated exciton can diffuse out of the focus in the case of the free rubrene surface. As, at 4 K, the decay time of the free exciton is on the order of 4 ns (see Fig. 3 (a)) this would require an exciton diffusion constant of at least 0.2 cm$^2$/s assuming an isotropic 2D diffusion within the focal plane. A comparison with temperature dependent diffusion constants of 10 cm$^2$/s, reported e.g. for anthracene single crystals at 4 K[8], indicates a reasonable agreement since we deduced a lower limit for this value. For the confined pyramidal structures one expects a shifting of the HOMO and LUMO positions towards higher energies at the terminating facet planes due to missing next neighbors and a reduced polarization energy compared to that of the bulk[14]. As a result, excitons approaching the facet surface experience a repelling potential at the boundary reflecting them back into the volume[15]. On average, the number of free excitons in the detection focus on the pyramids increases compared to that of the smooth surface. However, almost all self-trapped excitons remain in the detection volume since they are localized by lattice relaxation occurring on time scales of $10^{-12}$ to $10^{-11}$ s[7]. To further understand the intensity increase of the free exciton transition with the spatial extent of the pyramids the absorption characteristics of the respective structures have to be considered (see Fig. 2). As the rubrene absorption length for the excitation at $\lambda$= 380 nm amounts to 4 μm we conclude that for shallow pyramids, a substantial fraction of excitons is created in the bulk underneath and therefore not subjected to the confinement imposed by the microstructures. With increasing height, however, the absorption and thereby the generation of free excitons in the confined structures is continuously enhanced. By this model, the saturation of the 2.19 eV PL intensity for pyramids of at least similar height than the rubrene absorption length becomes clear.

Important questions raised by the PL spectra are: how does the conversion of the excited free exciton into the self-trapped state at elevated temperatures occur and what is the related energy barrier ? This phenomenon, which has been described in detail for several polyaromatic crystalline materials including rubrene[12,16–18], will also be considered here with a special focus on the impact of the confined surface geometry. The experimental decay times shown in Fig. 3(a) show no significant differences between the two excitonic species at 2.19 eV and 2.02 eV down to 30 K for different topographic positions (ON/OFF). The similarity between the exciton dynamics measured on and off the microstructures can be explained by the strong exciton-phonon interaction, by which the free exciton gains sufficient energy to surmount the barrier to the self-trapped state. As further indicated by the identical decay times, rate-limited occupation is sustained in this temperature range[17]. Below 30 K, a divergence of about 0.5 ns between the decay times of the two excitonic states can be detected. To characterize the underlying microscopic processes, we have fitted the experimental data by a theoretical description including a radiative decay and two non-radiative decay channels[19]:

$$\gamma(T) = \frac{1}{\tau(T)} = \gamma_r + \gamma_{nr1}e^{-\frac{E_1}{k_BT}} + \gamma_{nr2}e^{-\frac{E_2}{k_BT}} \quad (1)$$

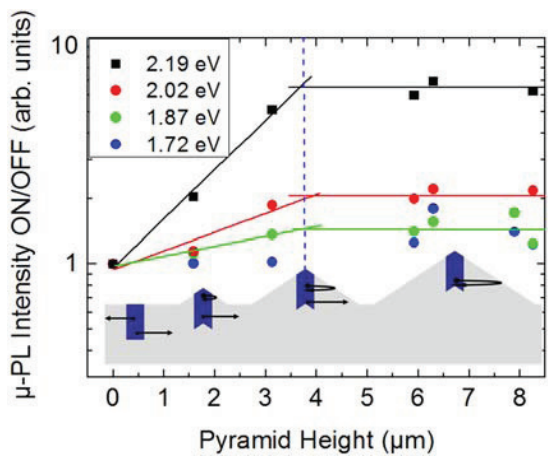

FIG. 2. Integrated PL intensity for various structure sizes normalized to that of the smooth rubrene surface. Pyramid heights above 4 μm result in almost constant intensity ratios. This height coincides with the rubrene absorption length at $\lambda$=380 nm as illustrated by the shaded profiles in the sketch.

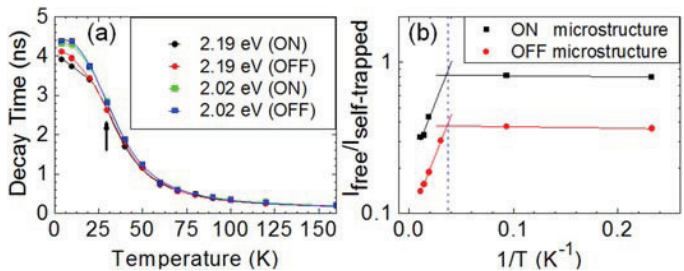

FIG. 3. (a) Temperature dependent decay times for the 2.19 eV and the 2.02 eV emission (dots). As indicated by the arrow, below 30 K a slight deviation in the decay times measured ON and OFF a microstructure occurs. The lines have been calculated according to Equation (1). (b) Temperature dependent intensity of the 2.19 eV emission normalized to the 2.02 eV emission including its vibronic replica. Below 0.04 $K^{-1}$ the non-radiative quenching yields an activation energy of 2 meV similar to $E_1$.

Here, $\gamma_r$ is the inverse lifetime of the radiative transition and $\gamma_{nri}$ describe the non-radiative quenching rates with activation energy $E_i$. The curves calculated accordingly (lines in Fig. 3 (a)) reveal activation energies for the non-radiative decay of $E_1 \approx 3$ meV and $E_2 \approx 14$ meV. These values are somewhat smaller than the respective activation energies for non-radiative decay of free and self-trapped excitonic states in the perylene compound PTCDA[19]. Comparing the adjusted rates $\gamma_r \approx 3 \cdot 10^8 s^{-1}$, $\gamma_{nr1} \approx 2 \cdot 10^8 s^{-1}$ and $\gamma_{nr2} \approx 10^{10} s^{-1}$ indicates that the second non-radiative mechanism is the dominant decay path. In this context, the high non-radiative decay rate $\gamma_{nr2}$ describes the homogeneous quenching of all PL channels resulting in halving of $\tau$ between 0 K and 30 K, whereas the low rate $\gamma_{nr1}$ is only responsible for the deviation occurring in this temperature regime. Due to the small energies $E_1$ and $E_2$ we attribute the related activation processes to the interaction between excitons and inter-molecular vibrations. Relating these energies to the lattice phonons observed by Raman measurements on rubrene single crystals[20], we find a correspondence of $E_2$ with the 104.8 $cm^{-1}$ or 118.6 $cm^{-1}$ mode and of $E_1$ with the lowest detected mode at 35.5 $cm^{-1}$.

Finally, we have analyzed the temperature-dependent ratio between the intensities of the two excitonic states (see Fig. 3 (b)). From this it becomes clear that for both positions (ON and OFF) the transition is thermally activated. The difference between the intensity ratios below 25 K (above 0.04 $K^{-1}$) is again proposed to originate from the diffusion of free excitons out of the detection focus on the plane surface compared to that under confined geometry. Above this temperature, both curves show a decrease in the area ratio. This is thought to evidence the efficient transfer of the free exciton into the self-trapped species. From the onset as well as from the linear slope we estimate an activation energy of 2 meV that is similar for the free surface and the confined geometry and confirms that estimated from the temperature dependent decay times.

In conclusion, we have investigated the optical properties of rubrene surface structures with confined geometries on the micrometer scale. We attribute the excitation to whether a free exciton or an exciton self-trapped by polarizing its molecular environment is formed. Although both geometries show similar decay times and activation barriers, for the confined geometry the fraction of free excitons turned out to be strongly enhanced with respect to that of the plane rubrene surface. This effect is explained by the respective optical absorption profile of the pyramids by which we identified a minimum height of about 4 $\mu$m to confine the generated excitons.

We acknowledge S. Hirschmann for assistance and the DFG for financial support (project FOR 730, SPP 1355)